\documentclass[12pt,sort&compress]{elsarticle}
\usepackage{amsmath}
\usepackage{amssymb}
\usepackage{amsthm}
\usepackage{bbm}
\usepackage{bm}
\usepackage{color}
\PassOptionsToPackage{hyphens}{url}
\usepackage[pdftex,colorlinks]{hyperref}
\usepackage[english]{babel}
\usepackage{natbib}
\newtheorem{lem}{Lemma}

\hypersetup{urlcolor=blue}
\def\Tr{\ensuremath{\mathrm{Tr}}}

\def\P{\ensuremath{\mathcal{P}}}
\def\Z{\ensuremath{\mathcal{Z}}}
\def\M{\ensuremath{\mathcal{M}}}
\def\W{\ensuremath{\mathcal{W}}}

\def\id{\ensuremath{\mathbbm{1}}}
\title{Measurement of a spin-1 system}
\author{Antonio Di Lorenzo}
\ead{dilorenzo@infis.ufu.br}
\address{Universidade Federal de Uberl\^{a}ndia, Uberl\^{a}ndia, MG, Brazil}
\address{CNR-IMM-UOS Catania (Universit\`a), Consiglio Nazionale delle Ricerche,
Via Santa Sofia 64, 95123 Catania, Italy}
\begin{document}
\begin{abstract}
We derive exact formulas describing an indirect measurement of a spin-1 system. 
The results hold for any interaction strength and for an arbitrary output variable $\hat{O}$. 
\end{abstract}
\maketitle
\section{Introduction}
The simplest non-trivial Hilbert space is the two-dimensional one, which describes a spin 1/2 or a qubit. 
The measurement of a spin 1/2 as realized in the Stern-Gerlach experiment\footnote{We remark that the Stern-Gerlach experiment provided the earliest direct evidence for the existence of spin, even though the hypothesis of spin would be advanced by Pauli only in 1924, two years after the experiment. Actually, Stern and Gerlach believed that the silver atoms had an angular momentum $L=1$, and their goal was to verify Bohr's prediction that the possible values of $L_z$ are quantized. The fact that the line corresponding to $L_z=0$ was missing from the experiment was overlooked. See {{\cite{Friedrich2003}}} for a recent historical account.} 
 \cite{Gerlach1922}
epitomizes the ideal quantum measurement, even though a realistic description of the measurement involves some complications \cite{Scully1987}. 

The next simplest system in quantum mechanics is provided by a three-dimensional Hilbert space, which can be realized, for instance, by a system with spin one. In the current jargon of quantum information, three-level systems are known as qutrits. They are known to provide higher-security quantum cryptography than qubits \cite{Bruss2002,Cerf2002}. Furthermore, it has been demonstrated that qutrits can be efficiently engineered and controlled, by using nonlinear optical techniques on bi--photons \cite{Bogdanov2004,Lanyon2008}. 
Spin--1 systems are important also for fundamental issues, as the Kochen-Specker theorem requires an Hilbert space at least three--dimensional \cite{Kochen1967}. 
In this context, the possibility of realizing an arbitrary projective measurement was questioned \cite{Hultgren1977} (we remark that a pure state $\psi$ of a spin 1/2 system is always an eigenstate of a spin component $\mathbf{n}\cdot\mathbf{S}$, so that any projective measurement reduces to the measurement of a spin--component, but the same does not hold for a spin--1 system). This challenge was answered positively \cite{Swift1980}. The validity of Kochen--Specker theorem for unsharp measurements on a spin--1 system was also 
questioned \cite{Pitowsky1985,Meyer1999,Kent1999,Clifton2000}, and it was shown that the theorem holds if the unsharpness is distributed covariantly \cite{Breuer2002}. 

Spin--1 systems are the only ones, besides the spin-1/2 systems, that satisfy a generalized idempotence relation $S^3=S$.  
To the best of my knowledge, there is no study of the general (i.e., non--projective) measurement of a spin-1, while a spin-1/2 has been treated 
quite extensively \cite{Scully1987,Peres1989,Duck1989,DiLorenzo2008}. 
In this manuscript, I am going to fill this gap, by studying a measurement of a spin--1 system followed, possibly, by a post--selection \cite{Aharonov1964}. 
General measurements, i.e., Positive-Operator Valued measures, are discussed in the books \cite{Davies1976,Helstrom1976,Kraus1983,Busch1995}; in particular, 
non--demolition measurements were treated in \cite{Braginsky1992}. Here, we shall consider linear non--demolition measurements, which means that the coupling between the 
detector and the system is linear in the measured operator $\Hat{S}$, and that the latter is conserved during the measurement process. 

In principle, for the special case of a detector having a continuous output, 
one could use the exact formal solution developed by Dressel and Jordan in \cite{Dressel2012c,Dressel2012d}, where the final density matrix of the system 
is expressed in terms of the initial density matrix and the initial Wigner function of the probe. 
However, these results apply only to the case when the readout variable of the detector is either canonically conjugated to or coincides with the variable appearing in the interaction, and the expression requires expanding the Wigner function in a series of its second argument, and then resumming, if possible, all the terms in the series. 
While for a spin-1 system it is possible to do so, as we show in the Appendix A, 
the procedure is unnecessarily complicated, and the more straightforward approach used here is better suited to the task. In Appendix B, we provide a slight improvement on the general formula of \cite{Dressel2012c}, by showing that it can be expressed in terms of the quantum characteristic function.

Finally, as our results apply to a detector having a discrete spectrum, they may be useful in Nuclear Magnetic Resonance implementations, where two nuclei, one with spin 1 the other with spin $S\ge 1$, interact. A recent realization of weak measurements (which is a limiting case of the results we present below) in NMR was reported in \cite{Lu2014}. 

%
\section{Background}

\subsection{A useful property of a spin-one operator}
In the following, we shall exploit the formula valid for a spin 1, 
\begin{equation}
\exp{( i \phi\hat{S})} = 1 +  i \sin{(\phi)}\, \hat{S} - [1-\cos{(\phi)}] \hat{S}^2 ,
\label{eq:expqt}
\end{equation}
which follows from 
\begin{equation}
\hat{S}^3=\hat{S}.
\label{eq:3=1}
\end{equation}
 We remark that this is the only property of the spin-one operator that we are going 
to exploit, so that the results presented here apply to any operator satisfying \eqref{eq:3=1}, not only operators 
on qutrits. In other words, the results of the present manuscript apply to any operator having eigenvalues in the set $\{-1,0,1\}$. 
Furthermore, the results can be trivially extended to any operator $\hat{X}$ having three equally spaced eigenvalues $x_1, x_2, x_3$, $x_2-x_1=x_3-x_2=\Delta x$, by making the shift and rescaling $\hat{X}=\Delta x \hat{S}+x_2$. 

In particular, an operator satisfying $\hat{S}^2=1$, e.g. a Pauli matrix representing a spin-1/2, satisfies also
~\eqref{eq:3=1}, so that the following results apply to this case as well, after applying the further restriction $\hat{S}^2=1$. 
As the exact solution of a measurement of a spin-1/2 is well known \cite{Peres1989,Duck1989,DiLorenzo2008,Kagami2011a,DiLorenzo2012a,Kofman2012}, it will provide a reference check. 
Furthermore, a projection operator satisfies as well~\eqref{eq:3=1}, but $\hat{S}^2=\hat{S}$.
Thus, the results presented in the following subsume both those for the measurement a spin-1/2 and those for 
the measurement of a \emph{yes/no} operator.

Another example of particular relevance where \eqref{eq:3=1} holds is that of two spin 1/2. 
Their total spin is a $4\times 4$ matrix, giving a reducible representation of SU(2). 
The sector corresponding to the singlet is represented by the scalar $0$, while the sector corresponding to the total 
spin 1 is represented by a $3\times 3$ operator $S_3$, namely
\begin{equation}
S = 
\left(
\begin{array}{cc}
0&\mathbf{0}_3^\dagger\\
\mathbf{0}_3&S_3
\end{array}
\right)
\end{equation}
with $\mathbf{0}_3$ the null vector in three dimensions. 

Recently, Aharonov \emph{et al.} have proposed to realize a quantum Cheshire cat \cite{Aharonov2013} 
by measuring the presence of a particle at a location, and its polarization at a separated location. 
In this case, in  the first location, a yes/no measurement is occurring, while in the second location a measurement 
of a local spin operator $\sigma$ is taking place. The latter operator can have the values $+1$ or $-1$ if the particle 
is there, and the value $0$ if the particle is not there. Therefore, the results presented in the following are relevant 
to extend the study of the quantum Cheshire cat to an arbitrary coupling \cite{DiLorenzo2014b}. 

\subsection{Description of the measurement}
In a measurement, before the interaction, the system and the detector are assumed to be uncorrelated, 
having a density matrix 
\begin{equation}
\rho^-=\rho_{i} \otimes \rho_{det};
\end{equation} 
the evolution operator of the system and the detector is taken to be the von Neumann interaction 
\begin{equation}
U = \exp( i \hat{Q}\hat{S}), 
\label{eq:timeevol}
\end{equation}
with $\hat{Q}$ an operator on the Hilbert space of the detector. 
The final entangled density matrix is thus 
\begin{equation}
\rho^+ = 
\exp( i \hat{Q}\hat{S}) \left(\rho_{i} \otimes \rho_{det}\right) \exp(- i \hat{Q}\hat{S}).
\label{eq:rhofin}
\end{equation}

We shall call the procedure a canonical measurement when the readout $\hat{P}$ has eigenstates 
$|j\rangle$ such that $\exp( i \hat{Q} S)$ translates one of them, say $|j_0\rangle$, into distinct eigenstates $|j_S\rangle$, with $S$ eigenvalues of the measured operator. 
Furthermore, we shall call the measurement ideal when the detector is prepared initially in the state 
$\rho_{det}=|j_0\rangle\langle j_0|$. 
A von Neumann measurement is an ideal canonical measurement. 
In the present manuscript, however, we shall consider measurements that obey \eqref{eq:timeevol}, and we shall not make the hypotheses of a canonical and ideal measurement, unless 
otherwise specified. Thus, we are using a von Neumann interaction, but we are dropping any further hypothesis behind the 
von Neumann model of measurement. 
In the case that the observable $\hat{P}$ of the detector is not canonically conjugated to $\hat{Q}$, the procedure 
could not be properly called a measurement, but perhaps an observation, in the sense that observing 
 $\hat{P}$ reveals something about the system, even though it is not a measurement of any observable $\hat{S}$. 
In particular, e.g., we could have $\hat{P}=\hat{Q}$, so that the variable does not change with the time-evolution operator 
$U=exp(i\hat{Q}\hat{S})$. In this case, observing $\hat{Q}$ does not yield information about $\hat{S}$, but about the  
``logarithmic directional derivative of the post--selection probability along the
flow generated by the unitary action of the operator $\hat{S}$'' \cite{Dressel2012d}.

\subsection{Post--selection}
The system may be post--selected in a state $E_{f}$, represented by a positive operator not necessarily having trace one \cite{Watanabe1955,Aharonov1964,Barnett2000,Barnett2001,Pegg2002}  by making a subsequent measurement. 
More precisely, the post--selected state is the normalized semipositive definite operator $E_{f}/\Tr(E_{f})$, which allows to make retrodictions about the 
past behavior of the system and which differs, in general, from the predictive state after the measurement $\rho_{f}$. Indeed, the two states coincide only if 
the post--selection measurement is a projective one. 
For instance, one could make a projective measurement of an observable 
$\hat{S}_{f}$, and analyze the output of the detector separately for each possible outcome $S_{f}$ \cite{Aharonov1964}. In this case, the post--selection 
states are the projectors $E_{f}=|S_{f}\rangle\langle S_{f}|$; or one could make a POV measurement of the system \cite{Wiseman2002}, then $E_{f}$ are not necessarily projectors; or, still, one could make a probabilistic post--selection of the data \cite{DiLorenzo2012a}. 

The reduced density matrix of the detector, for a given post--selection, is 
\begin{equation}
\rho_{det|f} = \frac{\Tr_{sys}[(E_{f}\otimes\id)\rho^+]}{\Tr_{sys,det}[(E_{f}\otimes\id) \rho^+]}
\label{eq:psdensmat}
\end{equation}
with $\Tr$ the trace, and $\Tr_{sys}$ the partial trace on the Hilbert space of the system. 
The normalization factor $\Tr_{sys,det}[(E_{f}\otimes\id)\rho^+]$ is the probability of successful post--selection $\P_{f}$.

%
\section{Results}
\subsection{General formula}
Usually, the output to be observed in the detector is $\hat{P}$, the variable conjugated to $\hat{Q}$. 
This implicitly requires that the detector has an infinite-dimensional Hilbert space, so that one can define canonically  
conjugated position and momentum operators. 
We shall not make this assumption and let, instead, the Hilbert space of the detector to be arbitrary. 

Let us start by computing the probability of post--selection. After substitution of \eqref{eq:expqt} into \eqref{eq:rhofin}, 
and expressing the trace over the detector Hilbert space in terms of position eigenstate, 
$\Tr_{det}[\dots] = \int dQ \langle Q|\dots|Q\rangle$, we have 
\begin{equation}
\P_{f} = \omega\left\{1-2\overline{\hat{s}} A''_{w} + \overline{\hat{s}^2} B_{w}
-2\overline{\hat{t}} C'_{w} +2\overline{\hat{s}\hat{t}} D''_{w} + 
\overline{\hat{t}^2} E_{w}\right\}, 
\label{eq:main0}
\end{equation}
where we defined 
\begin{subequations}
\begin{align}
\hat{s}=\sin{\hat{Q}},\\ 
\hat{t}=1-\cos{\hat{Q}}, 
\end{align}
\end{subequations}
the overline indicating average with respect to $\rho_{det}$, 
we introduced  
\begin{equation}
\omega = \Tr_{sys}[E_{f}\rho_{i}], 
\label{eq:om}
\end{equation}
the overlap between preparation and the post--selection, and we defined the weak values 
\begin{subequations}
\begin{align}
A_{w}= \omega^{-1} \Tr_{sys}[E_{f}\hat{S}\rho_{i}],
\label{eq:w1}
\\
B_{w}= \omega^{-1} \Tr_{sys}[E_{f}\hat{S}\rho_{i}\hat{S}],
\label{eq:w2}
\\
C_{w}= \omega^{-1} \Tr_{sys}[E_{f}\hat{S}^2\rho_{i}],
\label{eq:w3}
\\
D_{w}= \omega^{-1} \Tr_{sys}[E_{f}\hat{S}\rho_{i}\hat{S}^2],
\label{eq:w4}
\\
E_{w}= \omega^{-1} \Tr_{sys}[E_{f}\hat{S}^2\rho_{i}\hat{S}^2].
\label{eq:w5}
\end{align}
\label{eq:wv}
\end{subequations}
Notice that $B_{w}$ and $E_{w}$ are real, while $A_{w}$, $C_{w}$, and $D_{w}$ are complex. For brevity, we are indicating with a 
single prime the real part of a complex number, and with a double prime its imaginary part, $A_{w}=A'_{w}+iA''_{w}$, etc.
The quantities defined in \eqref{eq:wv} are called weak values just in analogy with the quantity defined in Ref.~\cite{Aharonov1988}, but we are not assuming anything here about the strength of the interaction. 

Without loss of generality, $\overline{\hat{O}}=0$, i.e. the average output before the interaction vanishes, which means 
that the detector is unbiased. Otherwise, if $\overline{\hat{O}}\neq 0$, one should substitute in the following 
$\delta\hat{O}=\hat{O}-\overline{\hat{O}}$ for $\hat{O}$. 
The average output is then 
\begin{align}
\langle O\rangle_{f}= \Tr_{det}[\hat{O}\rho_{det|f}]= \frac{\omega}{\P_{f}}\biggl\{
\overline{ i [\hat{O},\hat{s}]} A'_{w} - \overline{\{\hat{O},\hat{s}\}} A''_{w}
-\overline{\{\hat{O},\hat{t}\}} C_{w}'-\overline{ i [\hat{O},\hat{t}]} C''_{w} 
+\overline{\hat{s}\hat{O}\hat{s}} B_{w} 
\nonumber
\\
\qquad\qquad\qquad+
 i \,\overline{\hat{t}\hat{O}\hat{s}-\hat{s}\hat{O}\hat{t}} D'_{w}
-\overline{\hat{t}\hat{O}\hat{s}+\hat{s}\hat{O}\hat{t}} D''_{w}
+\overline{\hat{t}\hat{O}\hat{t}}  E_{w}\biggr\}.
\label{eq:main}
\end{align}

\subsection{Canonical continuous measurement}
In the following, we shall consider the case when $\hat{Q}$ has a continuous unbounded spectrum, so that it can be assimilated, 
say, to a position operator. The readout is taken to be its conjugated variable, $\hat{P}$.
For brevity, we shall overload the bar symbol with the following meaning: when applied to 
a function of $Q$ and $P$, with no hats, 
it represents quasi-averages, i.e. averages with respect to the initial Wigner function of the probe, namely 
\begin{equation}
\overline{f(P,Q)} = \int dP dQ W_{det}(P,Q) f(P,Q),
\label{eq:quasiav}
\end{equation}
with the Wigner function defined as 
\begin{equation}
W_{det}(P,Q)  
= \int \frac{dq}{2\pi}  e^{ i  qP} \langle Q-\frac{q}{2}| \rho_{det}|Q+\frac{q}{2}\rangle 
= \int \frac{dp}{2\pi}  e^{ i  Qp} \langle P+\frac{p}{2}| \rho_{det}|P-\frac{p}{2}\rangle .
\end{equation}
We note that when $f$ is a function only of $P$ or $Q$, then the quasi-averages are ordinary averages, 
$\overline{f(Q)}=\Tr[f(\hat{Q}) \rho_{det}]$ and 
$\overline{f(P)}=\Tr[f(\hat{P}) \rho_{det}]$. 
Then, the average output is 
\begin{align}
\langle P\rangle_{f}=  \frac{\omega}{\P_{f}}\biggl\{
\overline{\cos(Q)} A'_{w} - 2\overline{P\sin(Q)} A''_{w}
-2\overline{P[1-\cos(Q)]} C_{w}'-\overline{\sin(Q)} C''_{w} 
+\overline{P\sin^2{Q}} B_{w} 
\nonumber
\\
+
[1-\overline{\cos(Q)}] D'_{w}
+2\overline{P\sin(Q)[1-\cos(Q)]} D''_{w}
+\overline{P[1-\cos(Q)]^2}  E_{w}\biggr\}.
\label{eq:maincan}
\end{align}
This formula could be obtained in an alternative way by using Eqs. (3a), (5), and (7a) of Ref.~\cite{Dressel2012c}, after expanding the Eqs. (6) and (8a) therein, and then resumming the 
terms of the expansion. This procedure is implemented in Appendix A for the post--selection probability, while in Appendix B we provide a slight improvement on the results of  \cite{Dressel2012c} by showing that the expressions can be rewritten in terms of the Moyal quantum characteristic function of the detector, without the need for integration. 
\subsection{Canonical discrete measurement}
Here, we shall not assume that $\hat{Q}$ has a continuous spectrum. 
Let $d=2J+1\ge 3$ the dimension of the Hilbert space of the detector.
We shall assume that the readout $\hat{P}$ has eigenstates $|j\rangle$, $j\in\mathcal{I}=\{-J,-J+1,\dots,J\}$ 
and eigenvalues $P= j/\sqrt{d}$, such that 
$\exp[ i  k \hat{Q}]$ translates periodically them into each other, for any integer $k$ . Namely, 
\begin{equation}
\exp[ i  k\hat{Q}]|j\rangle = (-1)^{(d-1) r_{j+k}}|{j\oplus k}\rangle
\label{eq:modtrans}
\end{equation}
with $\oplus$ modular addition, i.e., the result of the ordinary sum $j+k$ is reduced to the interval $\mathcal{I}$ 
by adding or subtracting an appropriate multiple of $d$, $r_{j+k} d$.  
As discussed in Ref.~\cite{DiLorenzo2013f}, the two operators $\hat{Q}$ and $\hat{P}$ having this property can be considered 
a generalization of canonically conjugated operators in finite-dimensional Hilbert spaces. 
Actually, here we are abounding in requiring that \eqref{eq:modtrans} holds for all integer $k$. It would be sufficient, e.g.,  
that $\exp[i S\hat{Q}]|-J\rangle= |-J\oplus S\rangle$ for $S\in\{-1,0,+1\}$. 

We call this case the canonical discrete measurement because, if the initial state of the detector is 
$\rho_{det}=|-J\rangle\langle -J|$ and the system is in an eigenstate of $\hat{S}$, $\rho_{i}=|S\rangle\langle S|$, 
then the final state of the 
detector is one of the three orthogonal states $|-J\oplus S\rangle$, so that the von Neumann measurement criterion is satisfied \cite{vonNeumann1932}. 
However, as in the rest of this manuscript, we shall not make the further hypothesis that $\rho_{det}$ is an eigenprojector of $\hat{P}$. 

It follows that, in this case, the average readout is given by \eqref{eq:maincan}, as in the continuous case, 
but with the discrete Wigner function defined as 
\begin{equation}
W_{det}(P,Q) ={Re}[\langle \tilde{k}|j\rangle \langle j|\rho_{det}|\tilde{k}\rangle],
\label{eq:discrwig}
\end{equation}
with $|j\rangle$ the eigenstate of $\hat{P}$ corresponding to the eigenvalue $P= j/\sqrt{d}$ and 
$|\tilde{k}\rangle$ the eigenstate of $\hat{Q}$ corresponding to the eigenvalue $Q=2\pi k/\sqrt{d}$, 
\begin{equation}
|\tilde{k}\rangle =\frac{1}{\sqrt{d}}\sum_{j\in \mathcal{I}} \exp[-2\pi  i  j k/d] |j\rangle.
\end{equation}
Notice that the definition \eqref{eq:discrwig} ensures that the marginal probability obtained by either summing over $j$ 
or over $k$ is positive-definite 
\begin{align}
\sum_Q W_{det}(P,Q) =\sum_k {Re}[\langle \tilde{k}|j\rangle \langle j|\rho_{det}|\tilde{k}\rangle]= 
\langle j|\rho_{det}|j\rangle,\\
\sum_P W_{det}(P,Q) =\sum_j {Re}[\langle \tilde{k}|j\rangle \langle j|\rho_{det}|\tilde{k}\rangle]=
\langle \tilde{k}|\rho_{det}|\tilde{k}\rangle.
\end{align}
This property holds for any two bases $|j\rangle$, $|\tilde{j}\rangle$, not just for the canonically conjugated bases 
specifically considered here. For a review of discrete Wigner functions, see Ref.~\cite{Ferrie2011}. 
Interestingly, \eqref{eq:discrwig} is but the real part of the Kirkwood distribution function \cite{Kirkwood1933}. 
This distribution has been rediscovered and generalized several times in different contexts \cite{Terletsky1937,Dirac1945,Margenau1961,Rihaczek1968}, 
and its application to weak measurements has been pointed out \cite{Johansen2004a,Johansen2004b,Lundeen2012}. 

An important property used in deriving \eqref{eq:maincan} for the discrete case is that, even though 
the canonical commutation relation $[\hat{Q},\hat{P}]= i $ cannot be obeyed for finite $d$, however, the 
commutation relations $[\hat{P},\exp(\pm  i \hat{Q})]=\pm \exp(\pm  i \hat{Q})$ still hold, so that, e.g.,  
$[\hat{P},\sin(\hat{Q})]=- i \cos(\hat{Q})$, as if $\hat{P}=- i \partial/\partial Q$, formally.

\subsection{Spin 1/2 or $\hat{S}^2=1$}
We remark that, for a spin 1/2, or, more generally, for an operator satisfying $\hat{S}^2=1$, 
the following identities hold: 
$C_{w}=1$, $D_{w}=A_{w}$, $E_{w}=1$. Then, as expected, \eqref{eq:main} and \eqref{eq:main0} reduce 
to the expressions for a spin 1/2, as reported for instance in Refs.~\cite{DiLorenzo2012a,Kofman2012}. 
Furthermore, any operator $\hat{X}$ having only two distinct eigenvalues $x_1$ and $x_2$ can be reduced to this case by means 
of the transformation $\hat{X} =(x_1-x_2)\hat{S}/2+(x_1+x_2)/2$. 

\subsection{Yes-no measurement}
We remark that, for a yes--no measurement,$\hat{S}$ is a projection operator, i.e. $\hat{S}^2=\hat{S}$. 
Thus, the following identities hold: 
$C_{w}=A_{w}$, $D_{w}=B_{w}=E_{w}$.  

\subsection{Preparation or post--selection commuting with $\hat{S}$}
It may happen that either $[E_{f},\hat{S}]=0$ or $[\rho_{i},\hat{S}]=0$. 
Two important instances are when no post--selection is made, $E_{f}\propto \id$, and when the initial state is the 
completely unpolarized state $\rho_{i}=\id/3$. Other important cases are when the system is either prepared or post--selected in an eigenstate of $\hat{S}$. 
When this happens, all the weak values defined in \eqref{eq:wv} are real and furthermore 
$C_{w}=E_{w}=B_{w}$, $D_{w}=A_{w}$. 

\subsection{Preparation or post--selection commuting with $\hat{S}^2$}
It may happen that either $[E_{f},\hat{S}^2]=0$ or $[\rho_{i},\hat{S}^2]=0$, but $[E_{f},\hat{S}]\neq 0$ and 
$[\rho_{i},\hat{S}]\neq0$. While this case is less interesting than the former one, we shall treat it for completeness. 
The following relations hold among the weak values: 
$C_{w}$ is real, $D_{w}=A_{w}$, and $E_{w}=B_{w}$.

\subsection{Preparation and post--selection in pure states}
Since in this case $E_{f}\propto |f\rangle\langle f|$ and $\rho_{i}=|i\rangle\langle i|$, 
$B_{w}=|A_{w}|^2$, $E_{w}=|C_{w}|^2$, and $D_{w}=A_{w} C_{w}^*$.  
There are thus only four independent real parameters. 

\section{Weak measurement limit}
In this section, we shall compare the weak limit of our main result \eqref{eq:main0} and \eqref{eq:main} 
with the formulas for a weak measurement of any operator $\hat{S}$, 
which were given by Jozsa \cite{Jozsa2007} for the linear regime, and by me \cite{DiLorenzo2012a} in a more general case 
including orthogonal preparation and post--selection. 
As expected, the formulas coincide. 

In order to keep track of the perturbative expansion, it is better to introduce a coupling constant so that the time-evolution of \eqref{eq:timeevol} reads 
\begin{equation}
U = \exp( i \lambda\hat{Q}\hat{S}). 
\label{eq:timeevol2}
\end{equation}

\subsection{Conditions of validity for the weak measurement}
A measurement is called weak when $\lambda$ is sufficiently small. 
A sloppy way to characterize the strength of the measurement consists in saying that in the limit $\lambda\to\infty$ the measurement 
is strong, and in the limit $\lambda\to 0$ it is weak. However, $\lambda$ is a dimensionful constant (its dimension being the inverse of the dimension of $Q$, if we consider $S$ dimensionless), and it is common knowledge 
that dimensionful quantities are to be considered large or small always in comparison with another homogeneous quantity. 
Thus, the question to ask is: $\lambda$ is small compared to what? 
In the seminal paper of Aharonov \emph{et al.}, which considered a canonical continuous measurement, it 
was assumed that $\lambda \ll \sigma_P$, $\sigma_P^2$ being the initial variance in the canonical readout variable, i.e. the initial 
uncertainty over the pointer variable. 
However, as pointed out in Ref.~\cite{Duck1989}, the coherence of the detector, relative to the readout basis, is an 
essential requisite for the weak measurement to show its stranger features. Indeed, this was quantified better in Ref.~\cite{DiLorenzo2008}, 
where the initial state of the detector was considered to be a mixed Gaussian state 
\begin{equation}
\langle P|\rho_{det}|P'\rangle = \exp[-(P+P')^2/8\sigma_P^2-\sigma_Q^2(P-P')^2/2- i \overline{\hat{Q}}(P-P')],
\end{equation}
with $\sigma_Q^2$ the initial variance of the write-in variable $\hat{Q}$. 
Reference~\cite{DiLorenzo2008} showed that the relevant criterion for the weak measurement is that 
\begin{equation}
2\lambda \sigma_Q=\lambda/\delta P \ll 1,
\label{eq:wcond}
\end{equation} 
where $\delta P$ was defined as the coherence scale relative to the $|P\rangle$ basis, i.e. the scale over which the off--diagonal elements $\rho_{det}(P+p/2,P-p/2)$ vanish with respect to $\rho_{det}(P,P)$ for increasing $|p|$ and fixed $P$. 
Here and in the following, we assume that the range of the eigenvalues of $\hat{S}$ is $\mathcal{O}(1)$. If this were not the case, one could always redefine $\lambda$ and $\hat{S}$ appropriately. 
Since, by the Kennard uncertainty relation \cite{Kennard1927}, $\sigma_P\ge 1/2\sigma_Q=\delta P$, \eqref{eq:wcond} implies that 
$\lambda \ll \sigma_P$, but the vice versa may not be true. However, for a pure Gaussian state, $\sigma_P = 1/2\sigma_Q=\delta P$. 
Since this case was the one mostly considered in the literature, following Ref.~\cite{Aharonov1988}, the two scales $\delta P$ and $\sigma_P$ were not discriminated from each other, so that 
the more general condition for the weakness of the measurement, \eqref{eq:wcond}, was found only long after the concept of weak measurement had been established \footnote{Compare, however, Ref. \cite{Aharonov1990a}, where a similar condition 
was established in terms of the write--in variable, even though $\sigma_Q=1/2\sigma_P$ was assumed.
The condition \eqref{eq:wcond} was later rediscovered by Wu and Li \cite{Wu2011}, who were unaware of Ref.~\cite{DiLorenzo2008}.}.  
Furthermore, it is also required that $\lambda \overline{\hat{Q}}\ll 1$, but if this condition is not obeyed, one can gauge out 
$\overline{\hat{Q}}$ \cite{DiLorenzo2012a}. 
In Refs.~\cite{DiLorenzo2004,DiLorenzo2006}, the importance of the coherent behavior of the detector was stressed as well. 
There, however, the initial state of the detector was assumed to be that of an ideal von Neumann measurement, and the coherence 
was created by the very act of measuring the spin of several electrons sequentially. 
A more precise condition of validity for the perturbative expansion, including \eqref{eq:wcond} as a necessary condition, has been recently provided in Ref.~\cite{DiLorenzo2013g}, 
\begin{equation}
(2\lambda)^n \max\{|S|\}^n\overline{|\hat{Q}|^n}\le \delta^n , \forall n\in\mathbb{N}
\end{equation} 
with $\delta$ a small positive number. 

\subsection{Result}
By expanding the various terms in \eqref{eq:main0} and \eqref{eq:main} up to second order in $\lambda$
\begin{equation}
\P_{f} \simeq  \omega\left\{1-2\lambda\overline{\hat{Q}} A''_{w} + \lambda^2\overline{\hat{Q}^2} 
\left(B_{w}- C'_{w}\right) \right\}, 
\label{eq:main0w}
\end{equation}
and
\begin{equation}
 \langle O\rangle_{f}\simeq \frac{\omega}{\P_{f}}\biggl\{ 
\lambda\left(\overline{ i [\hat{O},\hat{Q}]} A'_{w} - \overline{\{\hat{O},\hat{Q}\}} A''_{w}\right)
+\frac{1}{2}\lambda^2\left(-\overline{\{\hat{O},\hat{Q}^2\}} C_{w}'-\overline{ i [\hat{O},\hat{Q}^2]} C''_{w} 
+2\overline{\hat{Q}\hat{O}\hat{Q}} B_{w} \right)
\biggr\},
\label{eq:mainw}
\end{equation}
in agreement with the result of Ref.~\cite{DiLorenzo2012a}. 
In particular, for $\hat{O}=\hat{Q}$ and $\hat{O}=\hat{P}$, the results of Ref.~\cite{Wu2011} are recovered.  
Furthermore, as noted in Refs.~\cite{DiLorenzo2012a} and \cite{DiLorenzo2012jj}, one can effectively neglect the $C_{w}$ terms, leading to the simplified interpolation formulas
\begin{equation}
\P_{f} \simeq  \omega\left\{1-2\lambda\overline{\hat{Q}} A''_{w} + \lambda^2\overline{\hat{Q}^2} 
B_{w} \right\}, 
\label{eq:main0w1}
\end{equation}
and
\begin{align}
\langle O\rangle_{f}\simeq  \frac{\omega}{\P_{f}}\biggl\{
\lambda\left(\overline{ i [\hat{O},\hat{Q}]} A'_{w} - \overline{\{\hat{O},\hat{Q}\}} A''_{w}\right)
+\lambda^2\overline{\hat{Q}\hat{O}\hat{Q}} B_{w} 
\biggr\}.
\label{eq:mainw1}
\end{align}
These formulas, originally derived and justified in Ref.~\cite{DiLorenzo2012a}, were also independently rediscovered by Kofman \emph{et al.} \cite{Kofman2012}. 

Finally, let us assume that it is admissible to make a Taylor expansion of $\omega/\P_{f}$, which is the case when $|A_{w}|$ and $B_{w}$ are not too large, i.e. when the preparation and the post--selection have not too small an overlap $\omega=\Tr(E_{f}\rho_{i})$. 
Then, we recover the formula due to Jozsa \cite{Jozsa2007},  
\begin{align}
\langle O\rangle_{f}\simeq   
\lambda\left(\overline{ i [\hat{O},\hat{Q}]} A'_{w} - \overline{\{\hat{O},\hat{Q}-\overline{\hat{Q}}\}} A''_{w}\right)
\label{eq:mainwlin}
\end{align}

%

\section*{Acknowledgments}
I am indebted to Luigi Amico for pointing out, several years ago, the simple property of a spin-1 which is at the base of the results presented herein. 
This work was performed as part of the Brazilian Instituto Nacional de Ci\^{e}ncia e
Tecnologia para a Informa\c{c}\~{a}o Qu\^{a}ntica (INCT--IQ) and 
it was supported by the Conselho Nacional de Desenvolvimento Cient\'{\i}fico e Tecnol\'{o}gico (CNPq) 
through process no. 245952/2012-8 and by Fundação de Amparo à Pesquisa de Minas Gerais (FAPEMIG) thorugh process no. PRI-00149-15. 

\appendix
%

\renewcommand*{\thesection}{\appendixname\ \Alph{section}}
\section{Alternative derivation for a canonical continuous measurement}
Here, we shall derive the main result following the approach of \cite{Dressel2012c,Dressel2012d} with a slight improvement that shows the connection with the classical and the quantum characteristic function of the detector \cite{DiLorenzo2013a}, rather than the partial Fourier transform of its Wigner function. 
We remark that this approach works only if $\Hat{Q}$, the write-in variable of the detector, has the continuous unbounded spectrum $(-\infty,+\infty)$, and is thus akin to a position operator. Notice that our variable $\Hat{Q}$ corresponds to $\hat{p}$ in \cite{Dressel2012c,Dressel2012d}, while the readout variable $\Hat{P}$ corresponds to $\hat{x}$. 

We shall use superoperators, i.e. operators that act on the linear space formed by the linear operators over the Hilbert space of the system. In particular, we shall need the adjoint action $\mathrm{ad}_{\Hat{S}}$ and the antiadjoint action $\mathrm{ac}_{\Hat{S}}$, defined by 
\begin{equation}
\mathrm{ad}_{\Hat{S}}\rho = \Hat{S}\rho-\rho\Hat{S},\quad 
\mathrm{ac}_{\Hat{S}}\rho = \Hat{S}\rho+\rho\Hat{S},\quad \forall \rho.
\end{equation}

Our starting point is the expression for the average conditional output $\langle P\rangle_f$
\begin{align}
\langle P\rangle &= \langle P\rangle_w + \langle S\rangle_w . 
\end{align}
Here, the expressions that we are going to use for the generalized weak values are equivalento to those of \cite{Dressel2012c}, the difference being that we are using the quantum characteristic function. Indeed, we write  
\begin{align}
\langle P\rangle_w =& \frac{-i\Tr_{sys}\left\{E_{f} \left.\dfrac{\partial}{\partial q}\M_{det}(\mathrm{ad}_{\Hat{S}},q)\right|_{q=0}\rho_i\right\}}{\P_{f}} ,
\label{eq:appcanav1}
\\
 \langle S\rangle_w =& \frac{\frac{1}{2}\Tr_{sys}\left\{E_{f} \mathrm{ac}_{\Hat{S}}\circ\M_{det}(\mathrm{ad}_{\Hat{S}},0)\rho_i\right\}}{\P_f} ,
\label{eq:appcanav2}
\end{align}
where $\P_{f}$ is the probability of post--selection, 
\begin{align}
\P_{f}& = \Tr_{sys}[E_{f} \M_{det}(\mathrm{ad}_{\Hat{S}},0)\rho_i] 
\label{eq:appmain0},
\end{align}
where $\M_{det}$ is the initial quantum characteristic function of the detector, defined in \eqref{eq:appmoyal}
These formulas are derived in Appendix B. 

First, we shall show how the expression for $\P_f$ \eqref{eq:appmain0} given here reduces to Eq.~\eqref{eq:main0}. 
We recall that the derivatives at the origin of the characteristic function are proportional to the average moments of $Q$, 
\begin{equation}
\left.\frac{\partial^n \M_{det}(p,0)}{\partial p^n}\right|_{p=0}=i^n \overline{Q^n} ,
\end{equation} 
where the average is taken with the initial state of the detector $\rho_{det}$. 
Therefore, Eq.~\eqref{eq:appmain0} has the Taylor expansion, 
\begin{align}
\P_{f}& = \sum_n \frac{i^n \overline{Q^n}}{n!} \Tr_{sys}[E_{f} \mathrm{ad}_{\Hat{S}}^n\rho_i] 
\label{eq:appmain1}.
\end{align}
We now prove the following lemma: 
\begin{lem}
For an operator satisfying $\Hat{S}^3=\Hat{S}$, the equation holds, 
\begin{align}
\mathrm{ad}_{\Hat{S}}^{2n+1} = \mathrm{ad}_{\Hat{S}} +(1-4^{n}) \mathrm{s}_{\Hat{S}}\circ \mathrm{ad}_{\Hat{S}}, \quad \forall n\in \mathbb{N},
\label{eq:applemma}
\end{align}
where the superoperator $\mathrm{s}_{\Hat{S}}$ is defined by
\begin{equation}
\mathrm{s}_{\Hat{S}}\rho = \Hat{S} \rho \Hat{S} .
\end{equation}
\end{lem}
(Since the associative property does not hold, the composition $\mathrm{s}_{\Hat{S}}\circ \mathrm{ad}_{\Hat{S}}$ here means that first we apply 
$ \mathrm{ad}_{\Hat{S}}$ to an operator 
of the Hilbert space, then the superoperator $\mathrm{s}_{\Hat{S}}$ to the result.) 
\begin{proof}
The proof follows by induction. After noting that \eqref{eq:applemma} holds trivially for $n=0$, while for $n=1$ we have 
\begin{align}
\mathrm{ad}_{\Hat{S}}^{3}\rho =& [\Hat{S},[\Hat{S},[\Hat{S},\rho]]] = \Hat{S}^3 \rho - 3\Hat{S}^2 \rho \Hat{S} + 3 \Hat{S} \rho \Hat{S}^2 - \rho \Hat{S}^3
\nonumber
\\
=& [\Hat{S},\rho]-3\Hat{S} [\Hat{S},\rho]\Hat{S} = \left(\mathrm{ad}_{\Hat{S}}-3 \mathrm{s}_{\Hat{S}}\circ \mathrm{ad}_{\Hat{S}}\right)\rho, \quad \forall \rho.
\end{align}
Therefore, supposing \eqref{eq:applemma} holds for $n-1$, we have 
\begin{align}
\mathrm{ad}_{\Hat{S}}^{2n+1} =& \mathrm{ad}_{\Hat{S}}^2 \circ \mathrm{ad}_{\Hat{S}}^{2n-1} =
\mathrm{ad}_{\Hat{S}}^3 
 +(1-4^{n-1}) \mathrm{ad}_{\Hat{S}}^2\circ \mathrm{s}_{\Hat{S}}\circ\mathrm{ad}_{\Hat{S}}
\nonumber\\
=&\mathrm{ad}_{\Hat{S}}-3 \mathrm{s}_{\Hat{S}}\circ\mathrm{ad}_{\Hat{S}}+(1-4^{n-1})\mathrm{s}_{\Hat{S}}\circ\mathrm{ad}_{\Hat{S}}^3
\nonumber
\\
=&\mathrm{ad}_{\Hat{S}}-3 \mathrm{s}_{\Hat{S}}\circ\mathrm{ad}_{\Hat{S}}+(1-4^{n-1})\mathrm{s}_{\Hat{S}}\circ\left(\mathrm{ad}_{\Hat{S}}-3\mathrm{s}_{\Hat{S}}\circ\mathrm{ad}_{\Hat{S}}\right)
\end{align}
where we used the fact that the operators $\mathrm{s}_{\Hat{S}}$ and $\mathrm{ad}_{\Hat{S}}$ commute. Now, since 
\begin{equation}
\mathrm{s}_{\Hat{S}}^2\circ\mathrm{ad}_{\Hat{S}}\rho=\Hat{S}^2 [\Hat{S},\rho]\Hat{S}^2 = \Hat{S}\rho\Hat{S}^2-\Hat{S}^2\rho\Hat{S} =
-\mathrm{s}_{\Hat{S}}\circ\mathrm{ad}_{\Hat{S}} \rho, \quad \forall \rho ,
\end{equation}
we have  
\begin{align}
\mathrm{ad}_{\Hat{S}}^{2n+1}=&\mathrm{ad}_{\Hat{S}}-3 \mathrm{s}_{\Hat{S}}\circ\mathrm{ad}_{\Hat{S}}+4(1-4^{n-1})\mathrm{s}_{\Hat{S}}\circ\mathrm{ad}_{\Hat{S}}
\nonumber
\\
=&\mathrm{ad}_{\Hat{S}}+(1-4^{n})\mathrm{s}_{\Hat{S}}\circ\mathrm{ad}_{\Hat{S}},
\end{align}
\end{proof}
For even powers, it follows immediately that 
\begin{align}
\mathrm{ad}_{\Hat{S}}^{2n+2} =& \mathrm{ad}_{\Hat{S}}^2 +(1-4^{n}) \mathrm{s}_{\Hat{S}}\circ \mathrm{ad}_{\Hat{S}}^2 
\nonumber
\\
=& 
\mathrm{ad}_{\Hat{S}}^2 +2 (1-4^{n}) \left(\mathrm{s}_{\Hat{S}} -\mathrm{s}_{\Hat{S}}^2\right),\quad \forall n\in\mathbb{N},
\label{eq:applemma2}
\end{align}
since 
\begin{equation}
\mathrm{s}_{\Hat{S}}\circ \mathrm{ad}_{\Hat{S}}^2 \rho =\Hat{S}\left(\Hat{S}^2\rho-2\Hat{S}\rho\Hat{S}+\rho\Hat{S}^2\right)\Hat{S}=2\left(\mathrm{s}_{\Hat{S}}-\mathrm{s}_{\Hat{S}}^2\right)\rho, \quad \forall \rho.
\end{equation}
By applying the lemma that we just proved to the Taylor expansion \eqref{eq:appmain1}, and separating the odd terms from the even terms in the sum we get 
\begin{align}
\P_{f}=&\Tr_{sys}[E_{f} \rho_i]+ \sum_{n=0} \frac{(-1)^n\overline{Q^{2n+1}}}{(2n+1)!} 
i\Tr_{sys}[E_{f} \mathrm{ad}_{\Hat{S}}\rho_i] 
\nonumber
\\
&+\sum_{n=0} \frac{(-1)^n(1-4^n)\overline{Q^{2n+1}}}{(2n+1)!} 
i\Tr_{sys}[E_{f} \mathrm{s}_{\Hat{S}}\circ\mathrm{ad}_{\Hat{S}}\rho_i] 
\nonumber
\\
&+\sum_{n=1} \frac{(-1)^n\overline{Q^{2n}}}{(2n)!} \Tr_{sys}[E_{f} \mathrm{ad}_{\Hat{S}}^2\rho_i] 
\nonumber
\\
&+2\sum_{n=1} \frac{(-1)^n(1-4^{n-1})\overline{Q^{2n}}}{(2n)!} \Tr_{sys}[E_{f} \mathrm{s}_{\Hat{S}}\rho_i] 
\nonumber
\\
&-2\sum_{n=1} \frac{(-1)^n(1-4^{n-1})\overline{Q^{2n}}}{(2n)!} 
\Tr_{sys}[E_{f} \mathrm{s}_{\Hat{S}}^2\rho_i] 
\label{eq:appmain2}.
\end{align}
Next, we note that the Taylor series can be resummed, yielding
\begin{subequations}
\begin{align}
\sum_{n=0} \frac{(-1)^n\overline{Q^{2n+1}}}{(2n+1)!} =&\ \overline{\sin(Q)},
\\
\sum_{n=0} \frac{(-1)^n(1-4^n)\overline{Q^{2n+1}}}{(2n+1)!}=&\ \overline{\sin(Q)[1-\cos(Q)]},
\\
\sum_{n=1} \frac{(-1)^n\overline{Q^{2n}}}{(2n)!}=&\ \overline{\cos(Q)}-1,
\\
\sum_{n=1} \frac{(-1)^n(1-4^{n-1})\overline{Q^{2n}}}{(2n)!}=&\ \overline{\cos(Q)}-1-\frac{1}{4}[\overline{\cos(2Q)}-1]
\nonumber
\\
=&\ -\frac{1}{2}\overline{[1-\cos(Q)]^2},
\end{align}
\end{subequations}
and furthermore that the traces over the system Hilbert space are 
\begin{subequations}
\begin{align}
\Tr_{sys}[E_{f} \rho_i]=&\ \omega,
\\
i\Tr_{sys}[E_{f} \mathrm{ad}_{\Hat{S}}\rho_i]=&\ -2\omega A''_w,
\\
i\Tr_{sys}[E_{f} \mathrm{s}_{\Hat{S}}\circ\mathrm{ad}_{\Hat{S}}\rho_i]=&\ 2\omega D''_w ,
\\
\Tr_{sys}[E_{f} \mathrm{ad}_{\Hat{S}}^2\rho_i]=&\ 2\omega (C'_w-B_w) ,
\\
\Tr_{sys}[E_{f} \mathrm{s}_{\Hat{S}}\rho_i]=&\ \omega B_w,
\\
\Tr_{sys}[E_{f} \mathrm{s}_{\Hat{S}}^2\rho_i] =&\ \omega E_w
.
\end{align}
\end{subequations}
These quantities were defined in \eqref{eq:om} and \eqref{eq:wv}. We recall that here $z'$ and $z''$ denote the real and imaginary part of $z$, respectively.  
Therefore, the first term on the right-hand side of \eqref{eq:appmain2} is but $\omega$ as defined in Eq.~\eqref{eq:om}. 
The second term is $-2\overline{\sin(Q)}\omega A''_w$. The third term is $2 \overline{\sin(Q)[1-\cos(Q)]}\omega D''_w$. 
The fourth term is  $2[1-\overline{\cos(Q)]}\omega (B_w-C'_w)$. The fifth term is $2[\overline{\cos(Q)}-3/4-\overline{\cos(2Q)}/4]\omega B_w$. 
Finally, the sixth term is $\overline{[1-\cos(Q)]^2}\omega E_w$. 
Adding up the terms, Eq.~\eqref{eq:main0} is recovered. 

The result \eqref{eq:maincan} may be derived by following the same procedure for the terms appearing in the numerator of 
\eqref{eq:appcanav1} and \eqref{eq:appcanav2}, after recalling that the quasi-averages \eqref{eq:quasiav} are generated by the quantum characteristic function, 
\begin{equation}
\overline{Q^m P^n} \equiv \int \!\!dP dQ\, W_{det}(P,Q) Q^m P^n = \left.\frac{(-i\partial)^{m+n}}{\partial p^m\partial q^n}\M_{det}(q,p)\right|_{q=p=0}.
\end{equation} 
Indeed, the numerator of $\langle P\rangle_w$, $-i\Tr_{sys}\left\{E_{f} \left.\dfrac{\partial}{\partial q}\M_{det}(\mathrm{ad}_{\Hat{S}},q)\right|_{q=0}\rho_i\right\}$, in \eqref{eq:appcanav1} differs from $\P_f$ only for the additional derivative in $q$, thus it is obtained by adding a $P$ variable to the quasi--averages. 
This procedure yields all the terms in the numerator of \eqref{eq:maincan} containing the variable $P$ in the quasi--averages.

The remaining terms are given by the numerator of $\langle S\rangle_w$ in \eqref{eq:appcanav2}, as we show in the following. 
We note that we have to apply half the anti--commutator superoperator to the terms in \eqref{eq:appmain2} inside the trace, yielding  
\begin{align}
&\Tr_{sys}\left\{E_{f} \mathrm{ac}_{\Hat{S}}\circ\M_{det}(\mathrm{ad}_{\Hat{S}},0)\rho_i\right\}	
=\ 
\nonumber\\
&\phantom{x+}\Tr_{sys}[E_{f} \mathrm{ac}_{\Hat{S}}\rho_i]+ \overline{\sin(Q)}
 i\Tr_{sys}[E_{f} \mathrm{ac}_{\Hat{S}}\circ\mathrm{ad}_{\Hat{S}}\rho_i] 
\nonumber
\\
&\phantom{x}+ \overline{\sin(Q)[1-\cos(Q)]} i\Tr_{sys}[E_{f}  \mathrm{ac}_{\Hat{S}}\circ\mathrm{s}_{\Hat{S}}\circ\mathrm{ad}_{\Hat{S}}\rho_i] 
\nonumber
\\
&\phantom{x}+[\overline{\cos(Q)-1]} \Tr_{sys}[E_{f}  \mathrm{ac}_{\Hat{S}}\circ\mathrm{ad}_{\Hat{S}}^2\rho_i] 
\nonumber
\\
&\phantom{x}+\overline{[1-\cos(Q)]^2} \Tr_{sys}[E_{f}  \mathrm{ac}_{\Hat{S}}\circ\mathrm{s}_{\Hat{S}}\rho_i] 
\nonumber
\\
&\phantom{x}-\overline{[1-\cos(Q)]^2} \Tr_{sys}[E_{f}  \mathrm{ac}_{\Hat{S}}\circ\mathrm{s}_{\Hat{S}}^2\rho_i] 
\label{eq:appmain3}.
\end{align}
Next, we note that the traces, recalling that $\Hat{S}^3=\Hat{S}$, are expressed in terms of 
\eqref{eq:om} and \eqref{eq:wv} as 
\begin{subequations}
\begin{align}
\Tr_{sys}[E_{f} \mathrm{ac}_{\Hat{S}}\rho_i]=&\ 2\omega A'_w,
\\
i\Tr_{sys}[E_{f} \mathrm{ac}_{\Hat{S}}\circ\mathrm{ad}_{\Hat{S}}\rho_i]=&\ -2\omega C''_w,
\\
i\Tr_{sys}[E_{f} \mathrm{ac}_{\Hat{S}}\circ\mathrm{s}_{\Hat{S}}\circ\mathrm{ad}_{\Hat{S}}\rho_i]=&\ 0,
\\
\Tr_{sys}[E_{f} \mathrm{ac}_{\Hat{S}}\circ\mathrm{ad}_{\Hat{S}}^2\rho_i]=&\ 2\omega (A'_w-D'_w) ,
\\
\Tr_{sys}[E_{f} \mathrm{ac}_{\Hat{S}}\circ\mathrm{s}_{\Hat{S}}\rho_i]=&\ 2\omega D'_w,
\\
\Tr_{sys}[E_{f} \mathrm{ac}_{\Hat{S}}\circ\mathrm{s}_{\Hat{S}}^2\rho_i] =&\ 2\omega D'_w
.
\end{align}
\end{subequations}
Adding up all the terms, we recover \eqref{eq:maincan}.
%
\section{Formulation of the conditional output in terms of the quantum characteristic function}
Here, we derive a formal solution for the canonical continuous measurement of an arbitrary operator $\Hat{S}$, following the approah of Ref.~\cite{Dressel2012c}. 
We provide a slight improvement, in the sense that our solution does not involve an integral over a Fourier transformed Wigner function, as per Eqs. (6) and (8a) of \cite{Dressel2012c}, but it is expressed in terms of the quantum characteristic function and its derivatives.  
We shall make use of the well known formula 
\begin{equation}
e^{\Hat{A}} \Hat{B} e^{-\Hat{A}} =\sum_{n=0}^{\infty} \frac{1}{n!}\overbrace{[\Hat{A},[\dots,[\Hat{A},\Hat{B}]\dots]]}^{n \text{ times}}=
e^{\mathrm{ad}_{\Hat{A}}} \Hat{B}. 
\end{equation}
The average value of the output variable $P$ conditional on a successful post--selection of the system in the final unnormalized state $E_f$ is obtained by averaging 
$\Hat{P}$ with the conditional state of the detector as given in Eq.~\eqref{eq:psdensmat}
\begin{equation}
\langle P\rangle_f = \frac{\Tr_{sys,det}[(E_{f}\otimes\Hat{P})U(\rho_i\otimes \rho_{det})U^\dagger]}{\Tr_{sys,det}[(E_{f}\otimes\id) U(\rho_i\otimes \rho_{det})U^\dagger]}
\end{equation}
Since we are assuming an interaction between the system and the probe giving rise to the evolution $U=\exp{(i\Hat{S}\Hat{Q})}$.
By using the cyclic property of the trace, and by writing 
\begin{equation}
E_{f}\otimes\Hat{P}=(E_{f}\otimes\id)U U^\dagger(\id\otimes\Hat{P}), 
\end{equation}
it follows that 
\begin{equation}
\langle P\rangle_f = \frac{\Tr_{sys,det}[U^\dagger(E_{f}\otimes\id) U (\Hat{P}+\Hat{S})(\rho_i\otimes \rho_{det})]}{\Tr_{sys,det}[U^\dagger(E_{f}\otimes\id) U(\rho_i\otimes \rho_{det})]},
\end{equation}
where we used the fact that $\Hat{Q}$ is the generator of the translations of $\Hat{P}$ (here, we are making use of the hypothesis that the two variables are canonically conjugated, hence the results in the following are limited to the case of a detector with a continuous output, as we mentioned above), so that 
\begin{equation}
U^\dagger\Hat{P}U = \Hat{P}+\Hat{S} .
\end{equation}
To be precise, we should write the right hand side as 
\begin{equation}
\id\otimes\Hat{P}+\Hat{S}\otimes\id , 
\end{equation}
but for brevity we use the same name to indicate the operators acting in the subspace of the system or of the detector, and the operators acting on the whole Hilbert space. 

Therefore, the conditional average becomes 
\begin{align}
\langle P\rangle_f &= \frac{\Tr_{sys,det}[U^\dagger(E_{f}\otimes\id) U \Hat{P}(\rho_i\otimes \rho_{det})]}{\Tr_{sys,det}[U^\dagger(E_{f}\otimes\id) U(\rho_i\otimes \rho_{det})]}
+\frac{\Tr_{sys,det}[(E_{f}\otimes\id) \Hat{S}U(\rho_i\otimes \rho_{det})U^\dagger]}{\Tr_{sys,det}[(E_{f}\otimes\id) U(\rho_i\otimes \rho_{det})U^\dagger]}
\nonumber
\\
&\equiv \frac{A(\Hat{P})}{A(\id)}+\frac{B(\Hat{S})}{B(\id)},
\label{eq:app0}
\end{align}
where we used the fact that $\Hat{S}$ commutes with $\Hat{U}$, and where we defined 
\begin{equation}
A(\Hat{P})= \Tr_{sys,det}[U^\dagger(E_{f}\otimes\id) U \Hat{P}(\rho_i\otimes \rho_{det})]
\end{equation}
and 
\begin{equation}
B(\Hat{S})= \Tr_{sys,det}[(E_{f}\otimes\id) \Hat{S} U (\rho_i\otimes \rho_{det})U^\dagger].
\end{equation}
Clearly, 
\begin{equation}
A(\id)=B(\id)=\Tr_{sys,det}[(E_{f}\otimes\id)  U (\rho_i\otimes \rho_{det})U^\dagger],
\end{equation}
the probability of a successful post--selection. 
Next, we proceed to the calculation of the functions $A$ and $B$. 
We use the $Q$ representation for the trace over the subspace of the detector, yielding 
\begin{align}
A(\Hat{P})&= \int\!\!dQ\, \Tr_{sys}[\exp{(-iQ\Hat{S})}E_{f} \exp{(iQ\Hat{S})}\rho_i] \langle Q| \Hat{P}\rho_{det}|Q\rangle, 
\\
B(\Hat{S})&= 
\int\!\!dQ\, \Tr_{sys}[E_{f}\Hat{S} \exp{(iQ\Hat{S})}\rho_i \exp{(-iQ\Hat{S})}] \langle Q| \rho_{det}|Q\rangle
\nonumber
\\
&=\Tr_{sys}[E_{f}\Hat{S} \int \!\!dQ \langle Q| \rho_{det}|Q\rangle
\exp{(iQ\mathrm{ad}_{\Hat{S}})}\rho_i] 
\nonumber
\\
&= \Tr_{sys}[E_{f}\Hat{S} \Z_{det}(\mathrm{ad}_{\Hat{S}})\rho_i] .
\end{align}
Here $\Z_{det}$ is the classical characteristic function relative to the statistics of the input variable $Q$, i.e., it is the Fourier transform 
of the probability $\P_{det}(Q)=\langle Q|\rho_{det}|Q\rangle$, 
\begin{equation}
\Z_{det}(p)\equiv \int \!\!dQ\, e^{ipQ} \P_{det}(Q).
\end{equation}
We recall the notion of quantum characteristic function, also called the Moyal function: it is the Fourier transform of the 
Wigner distribution, 
\begin{equation}
\M_{det}(p,q)\equiv \int \!\!dQ dP\, e^{ipQ+iqP} \W_{det}(Q,P).
\label{eq:appmoyal}
\end{equation}
Since the marginal of the Wigner quasi--probability, when the $P$ variable is ignored, is the probability $\P_{det}(Q)$, 
i.e., 
\begin{equation}
\int\!\ dP\, \W_{det}(Q,P) = \P_{det}(Q),
\end{equation}
it is immediate to verify that 
\begin{equation}
\Z_{det}(p)=\M_{det}(p,0) .
\end{equation}  
Therefore, we have that 
\begin{equation}
B(\Hat{S})= \Tr_{sys}[E_{f}\Hat{S} \M_{det}(\mathrm{ad}_{\Hat{S}},0)\rho_i] 
\end{equation}
and in particular 
\begin{equation}
B(\id)= \Tr_{sys}[E_{f} \M_{det}(\mathrm{ad}_{\Hat{S}},0)\rho_i] .
\end{equation}
The evaluation of $A(\Hat{P})$ is more involved:
\begin{align}
A(\Hat{P})&= 
\int\!\!dQ\, \Tr_{sys}\left[E_{f}\exp{(+iQ\Hat{S})}\rho_i \exp{(-iQ\Hat{S})}\right] \langle Q| \Hat{P} \rho_{det}|Q\rangle
\nonumber
\\
&=\Tr_{sys}\left[E_{f} \int\!\!dQ\,  \langle Q|\Hat{P} \rho_{det}|Q\rangle
e^{iQ\mathrm{ad}_{\Hat{S}}}\rho_i\right] 
\nonumber
\\
&=-i\Tr_{sys}\left[E_{f} \int\!\!dQ\,  \left.\frac{\partial}{\partial Q'}\rho_{det}(Q',Q)\right|_{Q'=Q}
\!\!\!e^{iQ\mathrm{ad}_{\Hat{S}}}\rho_i\right] 
\nonumber
\\
&=-i\Tr_{sys}\left[E_{f} \int\!\!dQ\!\left.\frac{\partial}{\partial Q'}\!\!\int\!\!dp\,e^{-ip(Q+Q')/2}\M_{det}(p,Q'-Q)\right|_{Q'=Q}
\!\!\!\!\!\!\!\!\!e^{iQ\mathrm{ad}_{\Hat{S}}}\rho_i\right] 
\nonumber
\\
&=-\frac{1}{2}\Tr_{sys}\left[E_{f} \int\!\!dQ\!\int\!\!dp pe^{-ipQ}\M_{det}(p,0)
e^{iQ\mathrm{ad}_{\Hat{S}}}\rho_i\right] 
\nonumber
\\
&\quad-i\Tr_{sys}\left[E_{f} \int\!\!dQ\!\!\int\!\!dp\,e^{-ipQ}\left.\frac{\partial}{\partial q}\M_{det}(p,q)\right|_{q=0}
e^{iQ\mathrm{ad}_{\Hat{S}}}\rho_i\right] . 
\end{align}
Finally, we have that 
\begin{equation}
A(\Hat{P})=
-\Tr_{sys}\left\{E_{f} \left[\frac{1}{2}\mathrm{ad}_{\Hat{S}}+\left.\frac{i\partial}{\partial q}\right]\M_{det}(\mathrm{ad}_{\Hat{S}},q)\right|_{q=0}\rho_i\right\} .
\end{equation}
Therefore, the numerator in the average conditional output \eqref{eq:app0} is 
\begin{align}
A(\Hat{P})+B(\Hat{S})&= 
\Tr_{sys}\left\{E_{f} \left[\frac{1}{2}\mathrm{ac}_{\Hat{S}}-\left.\frac{i\partial}{\partial q}\right]\M_{det}(\mathrm{ad}_{\Hat{S}},q)\right|_{q=0}\rho_i\right\} ,
\end{align}
where we introduced the superoperator defined by the anti--commutator, 
\begin{equation}
\mathrm{ac}_{\Hat{S}}=-\mathrm{ad}_{\Hat{S}}+2\Hat{S},\quad \mathrm{ac}_{\Hat{S}}\rho = \Hat{S}\rho+\rho\Hat{S},\quad \forall \rho.
\end{equation}

\end{document}